\documentclass{naturefig}
\usepackage{graphicx}

\newcommand{\rtwo}{r_{200}}

\newcommand{\msun}{M_\odot}

\newcommand{\vmax}{V_{\rm max}}

\newcommand{\rvmax}{r_{\rm Vmax}}
\newcommand{\vhost}{V_{\rm max,host}}
\newcommand{\ms}{\,{\rm m\,s^{-1}}}
\newcommand{\kms}{\,{\rm km\,s^{-1}}}
\newcommand{\sden}{\,{\rm M_\odot\,pc^{-3}}}
\newcommand{\psden}{\,{\rm M_\odot\,pc^{-3}\,km^{-3}\,s^{3}}}

\def\spose#1{\hbox to 0pt{#1\hss}}
\def\lta{\mathrel{\spose{\lower 3pt\hbox{$\mathchar"218$}} \raise 2.0pt\hbox{$\mathchar"13C$}}}
\def\gta{\mathrel{\spose{\lower 3pt\hbox{$\mathchar"218$}} \raise 2.0pt\hbox{$\mathchar"13E$}}}
\newcommand{\apj}{\textit{Astrophys. J.}}
\newcommand{\apjl}{\textit{Astrophys. J. Lett.}}
\newcommand{\apjs}{\textit{Astrophys. J. Supp.}}
\newcommand{\mnras}{\textit{Mon. Not. R. Astron. Soc.}}

\newcommand{\aap}{\textit{Astron. \& Astronphys.}}

\title{\Large \bf Clumps and streams in the local dark matter\\distribution}

\author{J. Diemand$^{1}$, M. Kuhlen$^2$, P. Madau$^1$, M. Zemp$^1$,
B. Moore$^3$, D. Potter$^3$, \& J. Stadel$^3$}

\begin{document}
\spacing{1}
\maketitle

\begin{affiliations}
 \item  University of California, Department of Astronomy and Astrophysics, 
Santa Cruz, CA 95064, USA.
 \item Institute for Advanced Study, Einstein Drive, Princeton, NJ 08540, USA.
 \item University Zurich, Institute for Theoretical Physics, Winterthurerstrasse 
190, 8057 Zurich, Switzerland.
 \end{affiliations}

\begin{abstract}
In cold dark matter cosmological models\cite{1982ApJ...263L...1P,1984Natur.311..517B},
structures form and grow by merging of smaller units\cite{1978MNRAS.183..341W}.
Numerical simulations have shown that such merging is incomplete; the inner cores of
halos survive and orbit as ``subhalos" within their
hosts\cite{1998MNRAS.300..146G,1999ApJ...522...82K}. Here we report a
simulation that resolves such substructure even in the
very inner regions of the Galactic halo. We find hundreds of very concentrated dark
matter clumps surviving near the solar circle, as well as numerous cold streams.
The simulation reveals the fractal nature of dark matter clustering:
Isolated halos and subhalos contain the same relative amount
of substructure and both have cuspy inner density profiles.
The inner mass and phase-space densities of subhalos match
those of recently discovered faint, dark matter-dominated
dwarf satellite galaxies\cite{2007arXiv0709.1510S,2007ApJ...670..313S,2007ApJ...654..897B},
and the overall amount of substructure can explain the anomalous flux ratios seen
in strong gravitational lenses\cite{2002ApJ...572...25D,2004ApJ...607...43M}.
Subhalos boost gamma-ray production from dark matter annihilation,
by factors of 4-15, relative to smooth galactic models. Local cosmic ray production
is also enhanced, typically by a factor 1.4, but by more than a factor of ten in one percent
of locations lying sufficiently close to a large subhalo. These estimates assume
that gravitational effects of baryons on dark matter substructure are small.
\end{abstract}

The cold dark matter (CDM) model has been remarkably successful at describing the large-scale mass distribution
of our Universe from the hot Big Bang to the present. However, the nature of the dark matter particle is best 
tested on small scales, where its interaction properties manifest themselves by modifying the 
structure of galaxy halos and their substructure. CDM theory predicts that the
growth of cosmic structures begins early, on Earth mass 
scales\cite{2004MNRAS.353L..23G,2005Natur.433..389D}, and continues
from the bottom up until galaxy clusters form that are twenty orders of magnitude more massive.
Resolving small-scale structures is extremely challenging,
as the range of lengths, masses, and timescales that need to be simulated is immense.
We have performed the highest precision calculation 
-- dubbed ``Via Lactea II'' -- of the assembly of the Galactic CDM halo.
The simulation follows the growth of a Milky Way-size system in a $\Lambda$CDM Universe from redshift 
104.3 to the present. It provides the most accurate predictions on the small scale clustering of dark matter 
and the first constraints on the local subhalo abundance and properties. We used the parallel treecode
PKDGRAV2[\cite{2001PhDT........21S}] and sample the galaxy-forming 
region with $1.1 \times 10^9$ particles of mass $4,100\,\msun$. 
Cosmological parameters were taken from {\it WMAP} 
[\cite{2007ApJS..170..377S}], see the online supplement for more details and a comparison to
our previous Via Lactea simulation.

The wealth of substructure that survives the hierarchical assembly process to the present epoch is clearly 
seen in Figure 1: we resolve over 40,000 subhalos within 
402 kpc of the center and they are distributed approximately 
with equal mass per decade of mass over the range $10^6-10^{9}\,\msun$. They have very high central 
phase-space densities ($\gta 10^{-5}\psden$) due to their steep inner density cusps and
their relatively small internal velocity dispersions. This agrees well with the 
extremely high phase-space densities inferred from stellar motions within
ultra faint dwarf galaxies\cite{2007ApJ...670..313S}. Our predicted inner subhalo densities
($0.4-2.5\,\msun$ pc$^{-3}$ within 100 pc, $7-46\,\msun$ pc$^{-3}$ within 10 pc)
are also in excellent agreement with the observations\cite{2007ApJ...670..313S,2007arXiv0709.1510S}.
The fact that CDM naturally predicts a small-scale dark matter distribution that matches
the observations is a real success of the model. Particle
candidates that introduce a low phase-space limit, such as a sterile neutrino, or that have a high 
collisional cross section such as self interacting dark matter would fail these fundamental observational 
tests.

The phase-space map (upper inset in Figure 1) also contains coherent elongated features.
These are streams which form out of material removed from accreted and disrupted subhalos.
The few visible streams have quite low densities 
(about 100 times below the local density) but owing to their low velocity dispersion
(about 10 times smaller than that of background particles) they just barely manage
to stand out in local phase space density (these streams have about $10^{-9}\psden$).
These resolved streams together with the multitude of expected finer grained phase space
structures that we currently can not resolve, will lead to unique signatures in direct detection
experiments, especially those with directional sensitivity.  
In cases where the disrupted subhalo hosted a luminous satellite galaxy, the resulting
streams would contain not only dark matter but also stars. This process would then
produce detectable features in the Milky Way's stellar halo, like those 
observed in the "Field of Streams"\cite{2006ApJ...642L.137B}.

Further evidence for halo substructure comes from the anomalous flux ratios in multiply-imaged 
gravitationally lensed quasars\cite{1998MNRAS.295..587M,2001ApJ...563....9M}. Perturbations to the light path 
from substructure can explain this phenomenon if the projected substructure fraction 
within 10 kpc is about one percent\cite{2002ApJ...572...25D,2004ApJ...607...43M}.
Within a projected distance of 10 kpc from the center, 0.50\% of 
the host mass belongs to resolved substructure, which could be just enough to explain 
the observed flux anomalies. In earlier simulations this fraction was lower,
even the first Via Lactea halo\cite{2007ApJ...667..859D} had only 0.25\% 
indicating that this quantity has not yet converged.

Via Lactea II predicts a remarkable self-similar pattern of clustering properties:
Our simulation is the first to use an extremely accurate integration of
particle orbits in high density regions\cite{2007MNRAS.376..273Z}, allowing
a precise determination of the density profile within
the inner kpc of the Galactic halo and within the centers of its
satellite galaxies. We find that a cuspy profile fits the host halos density profile well, while the best fit 
cored profile lies below the simulated inner densities (Figure 2).
The inner profiles of subhalos are also consistent with cusps over their resolved ranges.
They scatter around the moderate
cusp index of the host halo ($\gamma = 1.24$): Some of them are denser in the inner part, and some
are less dense, exactly like the inner parts of {\it field} halos, which
have inner slopes of $\gamma \simeq 1.2\pm0.2$[\cite{2004MNRAS.353..624D}].
At large radii subhalo density profiles generally fall off faster
than field halo profiles. These similarities and differences between
subhalo and field halo profiles have a simple explanation:
Subhalo density profiles were modified by tidal mass loss, which removes material
from the outside in, but does not change the
inner cusp structure\cite{2004ApJ...608..663K,2007ApJ...667..859D}.
Figure 3 shows that the dwarf satellites of the Milky Way 
appear to be scaled versions of the main halo not just in their inner mass distribution, but
also in term of relative substructure abundances.
Via Lactea II demonstrates the fractal-like appearance of the dark matter by resolving the second
generation of surviving sub-substructures from the merging hierarchy. This suggest that 
at infinite resolution one would find a long nested series of halos within halos within halos etc.,
reminiscent of a Russian Matryoshka doll, all the way down the first and smallest earth mass haloes that form.

The multitude of dark substructures increases the dark matter annihilation signal, since it is proportional to the
{\it square} of the local density. For cuspy profiles (Figure 2) with some fixed inner slope ($\gamma < 1.5$) one gets
the following simple scaling relation for the annihilation:
\begin{equation}
L \propto \rho_s^2 r_s^3 \propto \vmax^4/\rvmax \propto \vmax^3 \sqrt{c_V} \; ,
\end{equation}
see the online supplement for the definition of the concentration $c_V$ and its values.
Combined with the steep subhalo velocity function $N(>\vmax) \propto \vmax^{-3}$,
this implies that subhalos of all sizes contribute about
equally to the total signal coming from the Galactic dark halo. Taking the higher
concentrations of smaller systems\cite{2001MNRAS.321..559B}
into account, one finds that small subhalos are contributing more than large
ones\cite{2002PhRvD..66l3502U,2006AA...455...21C}. Summing up $\vmax^4/\rvmax$
of all resolved subhalos in Via Lactea II comes close (97\%) to the host halo's $\vmax^4/\rvmax$, i.e. the resolved subhalos
already contribute as much as their smooth host alone would.
In other words the "substructure boost factor" is at least 1.97.
Extrapolating down to micro-subhalos of size $\vmax=0.25 \ms$,
taking into account how concentrations depend on subhalo size\cite{2001MNRAS.321..559B}
and position (see the online supplement), and assuming a uniform distribution of subhalo
inner slopes $\alpha$ between 1.0 and 1.5, leads to a total boost of
14.6. Most of it comes from very small clumps: halting the same
extrapolation at $\vmax=44 \ms$ lowers the boost to 6.6.
While the contribution from small, dark clumps itself is not affected by
baryons, it may not dominate the total signal in scenarios in which
baryonic collapse greatly increases the central dark matter densities
in larger halos. However, the net effect of stars, black holes, and
galaxy formation is unclear, and it may actually lead to a reduction
in the central dark matter densities.
The detailed distribution of cusp indices is still unknown, since only
a few halos have been simulated with sufficient resolution\cite{2004MNRAS.353..624D}.
For the annihilation boost factors the existence of a few steep cusps near 1.5 would
make a big difference, since the signal diverges logarithmically towards the center in a $\gamma=1.5$ cusp.
Cutting the assumed uniform distribution of inner slopes at 1.4 instead of 1.5 gives
a boost of 9.9 instead of 14.6, and 4.3 instead of 6.6. These factors
imply that most of the extra-galactic $\gamma$-ray background
from dark matter annihilation\cite{2002PhRvD..66l3502U}, which will be constrained or even detected by the
upcoming GLAST mission, should be emitted by subhalos, and not by distinct host halos.

Besides $\gamma$-rays, dark matter annihilation would produce charged
particles and anti-particles that, due to to magnetic field entanglement,
propagate over much smaller distances within the Galaxy.
Space based experiments (like PAMELA, and in the near future
AMS-02) could detect anti-particles produced in dark matter
annihilations within about one kiloparsec\cite{2007arXiv0709.3634L}.
What fraction of this local annihilation would happen in nearby subhalos?
To constrain this local boost factor we use the same 
assumptions as above ($\gamma=[1 - 1.5]$,$\vmax \ge 0.25 \ms$), but now we only include subhalos
within one kiloparsec of the solar system (see the online supplement for the local subhalo abundance).
The resulting signal is 40\% of the smooth halo signal, giving a boost factor of 1.4, which we estimated using the
spherically averaged density at 8 kpc 
($\rho_0 = 0.40$ GeV c$^{-2}$ cm$^{-3}$). 
Explaining the positron excess from the HEAT experiment\cite{2004PhRvL..93x1102B} with
local dark matter annihilation requires enhancements from about 3 to 100\cite{2007arXiv0709.3634L}.
When a relatively large subhalo happens to lie within 1 kpc, one can get the higher boost factors
without violating the local subhalo constraints from our simulation. Such cases are possible, but 
not likely: Only 5.2 percent of all random realizations have a boost factor of 3 or larger (caused by a
$\vmax \ge 3.4 \kms$ clump within 1 kpc). In only 1.0 percent of the
cases the boost factor reaches 10 or higher due to a nearby, large $\vmax \ge 5.6 \kms$ subhalo.

\begin{addendum}
 \item[]Supplementary Information is linked to the online version of the paper at www.nature.com/nature.

 \item It is a pleasure to thank Bronson Messer and the Scientific Computing Group
at the National Center for Computational Sciences for their help. 
The ``Via Lactea II''  simulation was performed at the Oak 
Ridge National Laboratory through an award from DOE's Office of Science as part of
the 2007 Innovative and Novel Computational Impact on Theory and Experiment (INCITE) program.
Additional computations (initial conditions generation, code optimisations and smaller test runs)
were carried out on the MareNostrum supercomputer at the BSC, on Columbia at NASA Ames and
on the UCSC Astrophysics Supercomputer Pleiades. 
This work was supported by NASA and the Swiss National Science Foundation.

 \item[Competing Interests] The authors declare that they have no
competing financial interests.
 \item[Correspondence] Correspondence and requests for materials
should be addressed to J. D.~(email: diemand@ucolick.org).
\end{addendum}

\clearpage
\thispagestyle{empty}
\renewcommand{\figurename}{\bf Figure}
\begin{figure}
\begin{center}
\includegraphics[scale=1.0]{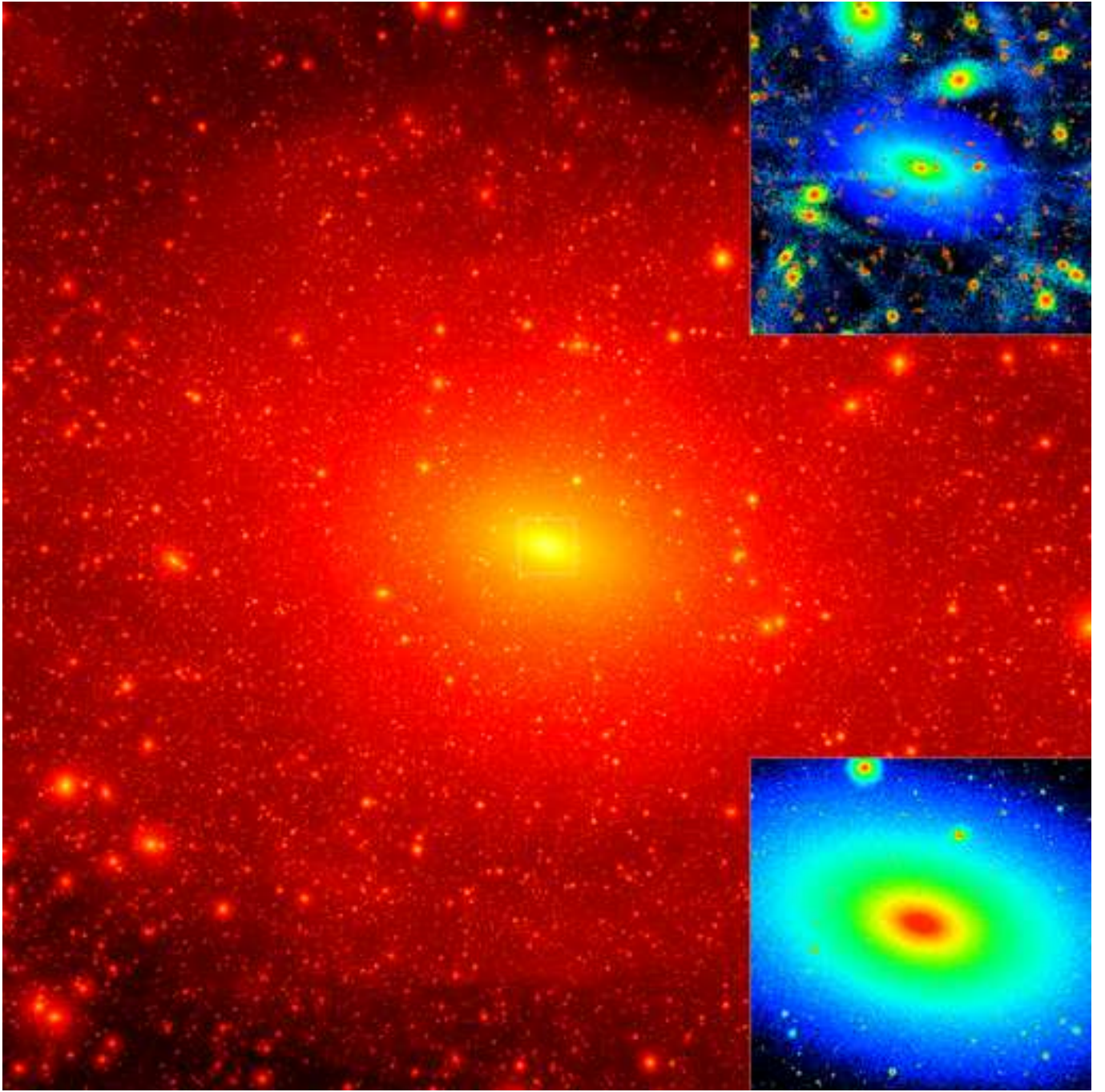}
\end{center}
\caption[]{\small Projected dark matter density-square map of ``Via Lactea II''. \rm
An 800 kpc cube is shown. The insets focus on an inner 40 kpc cube, in local density (bottom), 
and in local phase space density calculated with EnBiD[\cite{2006MNRAS.373.1293S}] (top).
The Via Lactea II simulation has a mass resolution of 4,100 $\msun$ and a 
force resolution of 40 pc. It used over a million processor hours on the ``Jaguar'' Cray XT3
supercomputer at the Oak Ridge National  Laboratory.
A new method was employed to assign physical, adaptive time-steps\cite{2007MNRAS.376..273Z}
equal to 1/16 of the local dynamical timescale (but not shorter than 268,000 yr),
which allows to resolve very high density regions.
Initial conditions were generated with a modified, parallel version of GRAFIC2[\cite{2001ApJS..137....1B}].
The high resolution region is embedded within a large periodic box
(40 comoving Mpc) to account for the large scale tidal forces.
The mass within 402 kpc (the radius enclosing 200 times the mean matter density) 
is $1.9\times 10^{12}\,\msun$. 
}
\label{fig1}
\end{figure}

\clearpage
\thispagestyle{empty}
\begin{figure}
\begin{center}
\includegraphics[scale=0.75]{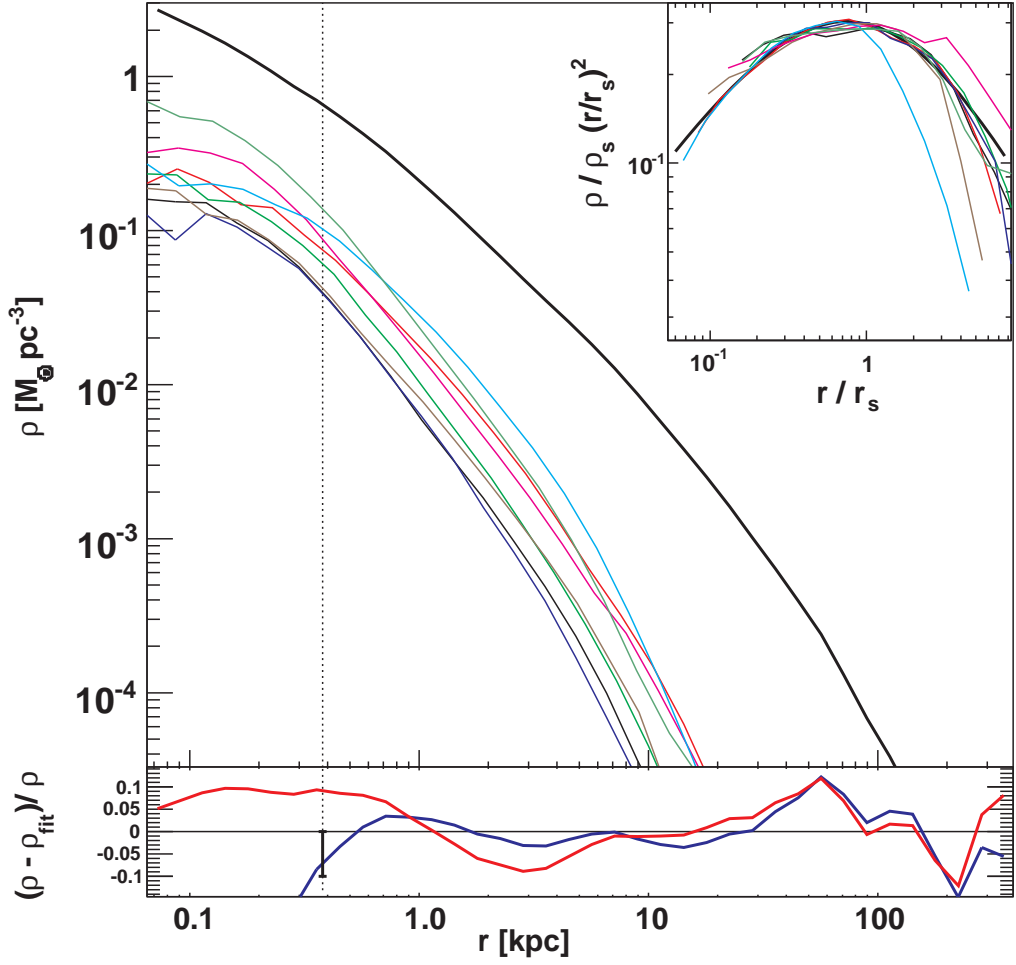}
\end{center}
\caption[]{Density profiles of main halo and subhalos. \rm
\small {\it Main panel}: Profile of the Milky Way
halo ({\it thick line}) and of eight large subhalos ({\it thin lines}).
The {\it lower panel} gives the relative differences between
the simulated main halo profile and a fitting formula 
with a core\cite{2004MNRAS.349.1039N}
$\rho(r) = \rho_s  \exp \{ -2/\alpha \left[ (r/r_{s})^{\alpha} - 1 \right]$,
with best fit parameters: $\alpha =0.170$, $r_s = 21.5\, {\rm kpc}$,     
$\rho_s=1.73 \times 10^{-3} \sden$
({\it red curve}) and one with a 
cusp\cite{2004MNRAS.353..624D} $\rho(r) = \rho_s (r/r_{s})^{-\gamma} (r/r_{s} + 1)^ {-3+\gamma}$
with a best fit inner slope of $\gamma = 1.24$,
$r_s = 28.1\, {\rm kpc}$, $\rho_s=3.50 \times 10^{-3} \sden$ ({\it blue curve}).
The vertical dotted line indicates the
estimated convergence radius of 380 pc: simulated local densities
are only lower limits inside of 380 pc and they
should be correct to within 10\% outside this region.
The cuspy profile is a good fit to the inner halo, while the cored profile
has a too shallow slope in the inner few kpc, causing it to overestimate
densities around 4 kpc and to underestimate them
at all radii smaller than 1 kpc. The simulated densities
are higher than the best fit cored profile even at 80 pc, where they are certainly
underestimated due to numerical limitations.
We find the same behavior in the inner few kpc in all six snapshots we have analyzed so far  
between z=3 an z=0. The large residuals in the outer halos on the other hand are transient
features, they are different in every snapshot.
{\it Inset}: Rescaled host (thick line) and subhalo (thin lines) density profiles multiplied
by radius square to reduce the vertical range of the figure. 
}
\label{fig:profile}
\end{figure}

\clearpage
\thispagestyle{empty}
\begin{figure}
\begin{center}
\includegraphics[scale=0.75]{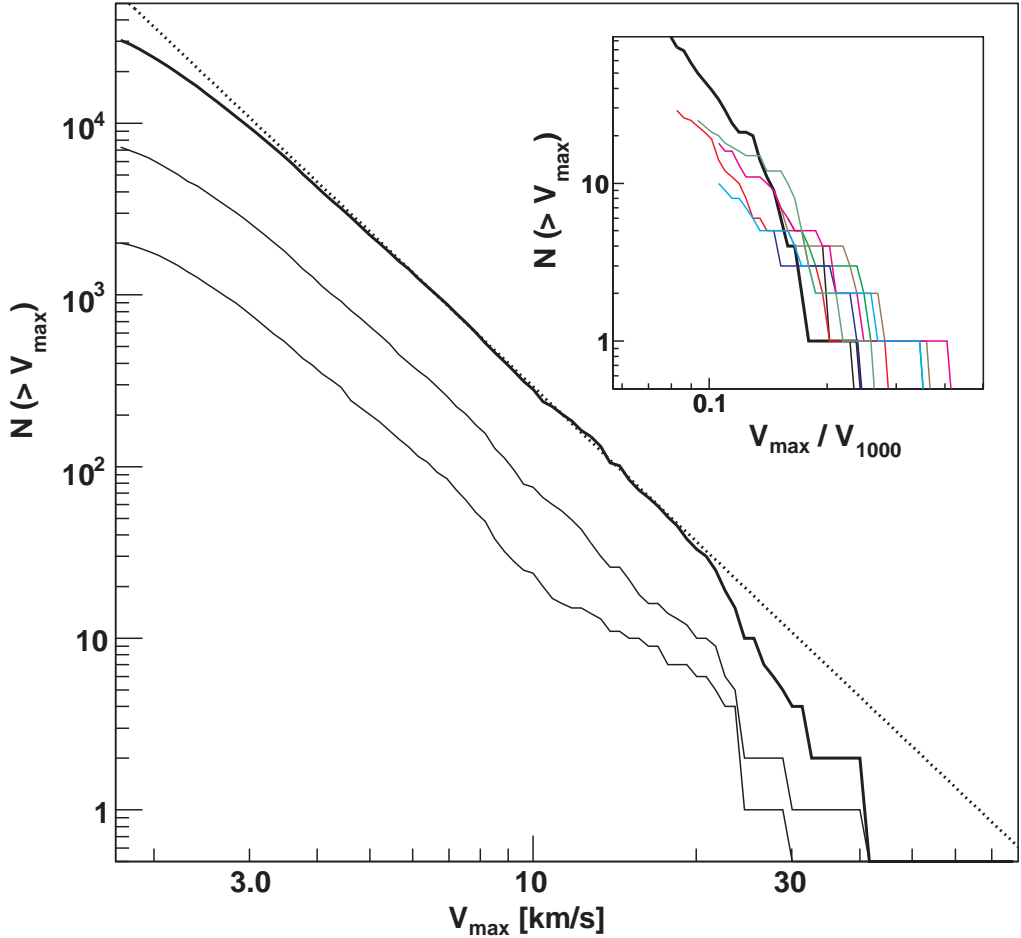}
\end{center}
\caption{Subhalo and sub-subhalo abundances.  \rm Number of subhalos above
$\vmax$ within $\rtwo=402$ kpc (thick solid lines) and within 100 and 50 kpc of the
galactic center (thin solid lines). $\vmax$ is the peak height of the 
subhalo circular velocity $v_{circ} = \sqrt{GM(<r) / r}$ and serves as
a simple proxy for the mass of a subhalo. The dotted line is
$N(>\vmax)=0.036\,(\vmax / V_{\rm max,host})^{-3}$, where $V_{\rm max,host} = 201 \kms$
(at $r_{\rm Vmax,host}= 60$ kpc).
It fits the subhalo abundance above $\vmax \simeq 3.5 \kms$.
The number of smaller subhalos is artificially reduced by numerical
limitations. Inside $\rtwo$ this halo has 1.7 times more substructure than
the first Via Lactea halo\cite{2007ApJ...667..859D}, a factor well within
the halo-to-halo scatter\cite{2005MNRAS.359.1537R}. Inside 50 kpc the difference
grows to 2.6, probably due to the improved mass and time resolution
of Via Lactea II, which allows to resolve inner substructure better. 
The {\it inset} shows the sub-subhalo abundance within $r_{1000}$ 
(enclosing 1000 times the mean matter density) of the centers
of eight (same ones as in Fig. 2) large subhalos (thin solid lines). $r_{1000}$
is well inside of the tidal radius for these systems. The thick solid line
shows the subhalo abundace of the host halo inside of its $r_{1000} = 213$ kpc.
The (sub-)subhalo $\vmax$ values are given in
units of $V_{1000}=\sqrt{GM(<r_{1000}) / r_{1000} }$
of the corresponding host (sub-)halo. Lines stop at $\vmax = 2 \kms$.
The mean sub-substructure abundance is consistent with 
the scaled down version of main halo, and both the mean abundance and the scatter agree
with results in\cite{2005MNRAS.359.1537R} for distinct {\it field} halos.
}
\label{fig3}
\end{figure}
\clearpage


\spacing{1}
{\large\bfseries\noindent\sloppy \textsf{Supplementary Notes}}

In this Supplementary Information, we give more details about the Via Lactea II (VL-II) simulation presented in 
our Letter. The first part provides additional results about radial trends in subhalo abundance and properties. 
In the second part we include some comparisons with the Via Lactea (VL-I) run, previously the largest simulation 
of the Galactic dark matter halo.

Supplementary Figure 1 illustrates how tidal interactions with the gravitational potential of the host 
shape the radial distribution of subhalos and their internal structure. Tides remove mass from the outer 
parts of subhalos. Clumps closer to the Galactic center lose more mass as they experience stronger tidal 
forces and more pericenter passages$^{18}$. Mass loss also reduces $\vmax$ and $\rvmax$, which leads 
to an increase in $\bar{\rho}(<r_{\rm Vmax})$. The radial distribution of subhalos with $\vmax > 3 \kms$ 
is more extended than the dark matter distribution in the Galactic halo, a feature that does not depend
on subhalo size, i.e. different $\vmax$ selection thresholds lead to the same radial distribution. 
VL-II resolves subhalos as close as 8 kpc from the Galactic center, but it is possible that it underestimates the true 
substructure abundance inside 20 kpc due to numerical limitations. Subhalo concentrations are defined as 
the mean density within $\rvmax$, the radius of peak circular velocity, a quantity that is well 
determined both for isolated halos and subhalos and does not dependent on assumptions 
about their density profiles$^{18}$:
\begin{equation}
c_{V} \equiv  \bar{\rho}(<r_{\rm Vmax}) /  \rho_{\rm crit,0} 
=  \left( V_{\rm max} / r_{\rm Vmax} \right)^2 4\pi/ (3 G \rho_{\rm crit,0}),
\end{equation}
where $\rho_{\rm crit,0}=1.48 \times 10^{-7}\,\sden$. The median subhalo concentration increases
strongly towards the Galactic center, both because of tidal mass losses and to a lesser extent because
of the earlier formation times of inner substructure$^{18}$.

Supplementary Table 1 summarises the numerical parameters and host halo properties of the VL-I and 
VL-II simulations. The main difference between these runs is the improved time stepping in VL-II, 
which at each time and for each particle is based on the local dynamical time 
$\sqrt{1/G\rho_{\rm enc}}$$[^{19}]$, where $\rho_{\rm enc}$ is the mean
density enclosed in a sphere extending out to the particle's position and centered on the 
dynamically dominant structure. In VL-I we adopted instead the standard ad-hoc time-step 
criterion based on the acceleration and the gravitational softening of the particle ($\propto 
\sqrt{\epsilon/|{\bf a}|}$). This criterion fails in high density regions and artificially flattens
the inner cuspy halo density profiles$^{31}$. This limitation set the convergence 
radius of VL-I (where the host true density profile is reproduced to within 10 percent) to
$r_{\rm conv}=1.3\,$ kpc. The new time stepping used in VL-II allows us to properly resolve 
the host density profile on significantly smaller scales: the finite mass resolution determines 
a convergence radius$^{20}$ of about 0.38 kpc for VL-II. The VL-I subhalo density profiles are also
affected by this limitation: the enclosed densities within 300 pc of large subhalos are about twice 
as high in VL-II. Further out, at 600 pc, the enclosed subhalo densities are very similar in VL-I and VL-II.

To check for numerical convergence and test the dependence of our results on numerical and 
cosmological parameters we ran a series of lower resolution versions of VL-II. The mass resolution in 
this "VL-IIm" series is 64 times coarser and the force softening length 4 times larger than in VL-II. 
Supplementary Figure 2 shows that the lower resolution version of the same halo has a very similar 
subhalo velocity function above about 0.05 $\vhost \simeq 10 \kms$. Rescaling to 64 times
less massive systems suggests that VL-II should 
have converged down to about 2.5 $\kms$, which indeed is close to the scale where the VL-II velocity 
function starts to fall below the power law fit (Figure 3). The earlier starting redshift of VL-II 
does not seem to affect the $z=0$ substructure abundance significantly. We find only a weak 
dependence on cosmological parameters: VL-II used the best fit $\Lambda$CDM parameters from the 
{\it WMAP} 3 year data release$^{14}$:
$\Omega_m =$ 0.238, $\Omega_{\Lambda} =$ 0.762, h=0.73, $n_s$=0.951, and $\sigma_8$=0.74. For comparison we
have ran a simulation with {\it WMAP} 1 year parameters$^{32}$:
$\Omega_m =$ 0.27, $\Omega_{\Lambda} =$ 0.73, h=0.72, $n_s$=1.0, and $\sigma_8$=0.9. The 
higher $\sigma_8$ and steeper spectral index $n_s$ in {\it WMAP1} lead to more
small scale power. The effect on the $z=0$ substructure abundance 
is rather small: our WMAP1 run has about 20 to 30 percent more subhalos relative to the WMAP3 runs,
in agreement with semi-analytical predictions (see Figure 11 in$^{33}$).


31. Diemand, J., Zemp, M., Stadel, J., Moore, B. $\&$ Carollo, C. M. Cusps in cold dark matter
haloes. Mon. Not. R. Astron. Soc. 364, 665--673 (2005). 

32. Spergel, D. N. et al. First-Year Wilkinson Microwave Anisotropy Probe (WMAP)
Observations: Determination of Cosmological Parameters. Astrophys. J. Supp. 148, 175--194 (2003). 

33. Zentner, A. R. $\&$ Bullock, J. S. Halo Substructure and the Power Spectrum. Astrophys. J. 598, 49--72 (2003). 


\renewcommand{\tablename}{\bf Supplementary Table}
\begin{table}
\begin{center}
\spacing{1.5}
\begin{tabular}{lcccccccccc}
\hline \hline
Name & $\Delta T$ &$\epsilon$ & $z_i$ &$r_{\rm conv}$& $M_{\rm hires}$ & $r_{200}$ & $M_{200}$ & $\vmax$       & $\rvmax$  \\
     & & (pc)       &       & (kpc)     &      ($\msun$)       & (kpc)     & ($\msun$) & (km s$^{-1}$) & (kpc)               \\
\hline
VL-I&$0.2\sqrt{\epsilon/|{\bf a}|}$& 90.0 & 48.4 & 1.3 & $2.1 \times 10^4$ & 389 & $1.77 \times 10^{12}$ & 181 & 69 \\
VL-II &$0.06\sqrt{1/G\rho_{\rm enc}}$& 40.0 & 104.3 & 0.38 & $4.1 \times 10^3$ & 402 & $1.93 \times 10^{12}$ & 201 & 60 \\
\hline
\end{tabular}
\spacing{1}
\setcounter{table}{0}
\caption[]{Simulation parameters and halo properties.
  \rm Time step criterion $\Delta T$, spline force softening length
  $\epsilon$, initial redshift $z_i$, convergence radius of the host halo
  density profile $r_{\rm conv}$, mass
  $M_{\rm hires}$ of high resolution dark matter particles and host halo
  $r_{200}$, $M_{200}$, $\vmax$ and $\rvmax$ for the VL-I and VL-II
  simulations. At each time individual particle time steps are chosen by dividing the base
  time step of 13.7 Gyr / 400 by two until it is smaller than $\Delta T$,
  where $|a|$ is the norm of the acceleration vector and 
  $\rho_{\rm enc}$ is the enclosed density within the dynamically dominant structure.
  Force softening lengths $\epsilon$ are constant in
  physical units back to $z=9$ and constant in comoving units
  before.}
\label{table:simulations}
\end{center}
\end{table}

\renewcommand{\figurename}{\bf Supplementary Figure}

\begin{figure}
\begin{center}
\includegraphics[scale=0.65]{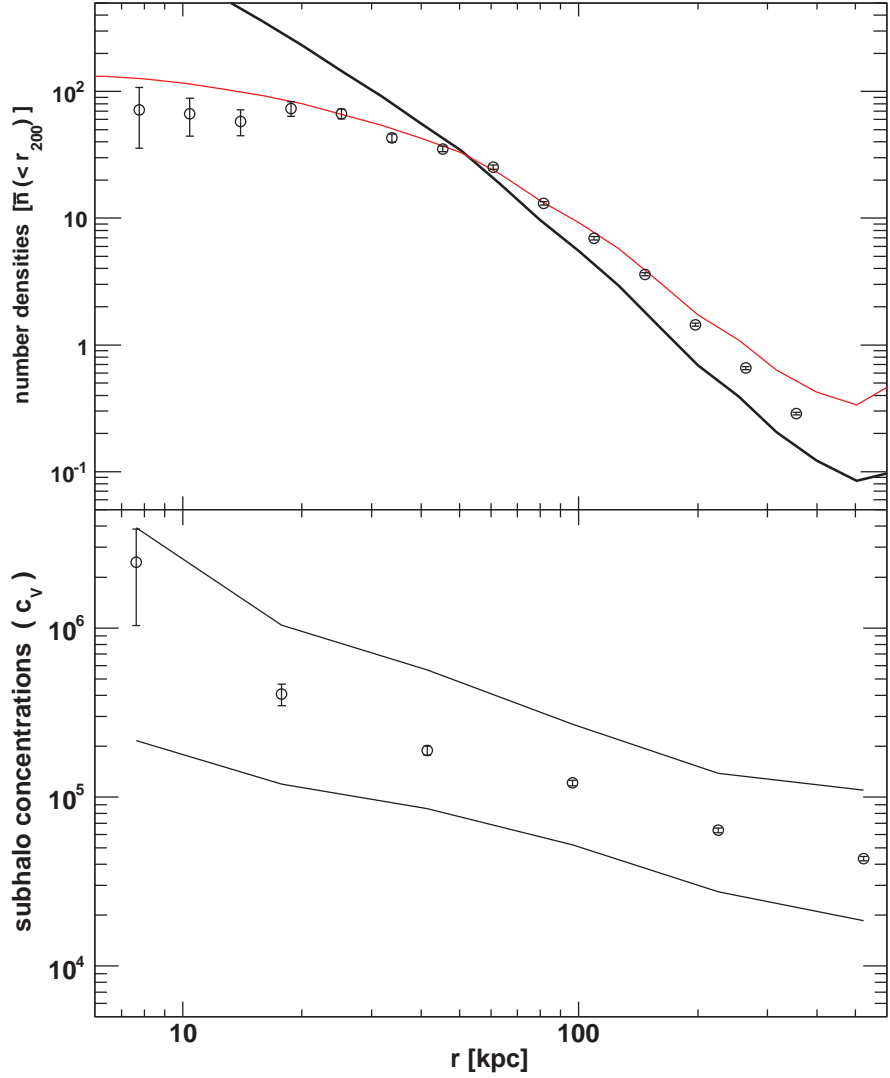}
\end{center}
\caption[]{Abundance and concentrations of subhalos vs. distance from the galactic center. \rm
{\it Top:} The number density profile of subhalos (circles) is more extended than  the dark matter
density profile $\rho(r)$ (thick line). Their ratio turns out to be roughly proportional to the enclosed mass $M(<r)$,
i.e. $\rho M(<r)$ (thin line) matches the subhalo number density quite well.
Only subhalos larger than $\vmax = 3 \kms$ are included here.
{\it Bottom:} Subhalo concentrations (median and 68\% range are shown) 
increase towards the center, where the stronger tidal force remove more of the outer, low density
parts from the subhalos. 
To make sure their $c_{V}$ are resolved, only subhalos larger than $\vmax = 5 \kms$ are used.
The error bars indicate the statistical uncertainties in both panels.
}
\label{fig:radial}
\end{figure}

\begin{figure}
\begin{center}
\includegraphics[scale=0.7]{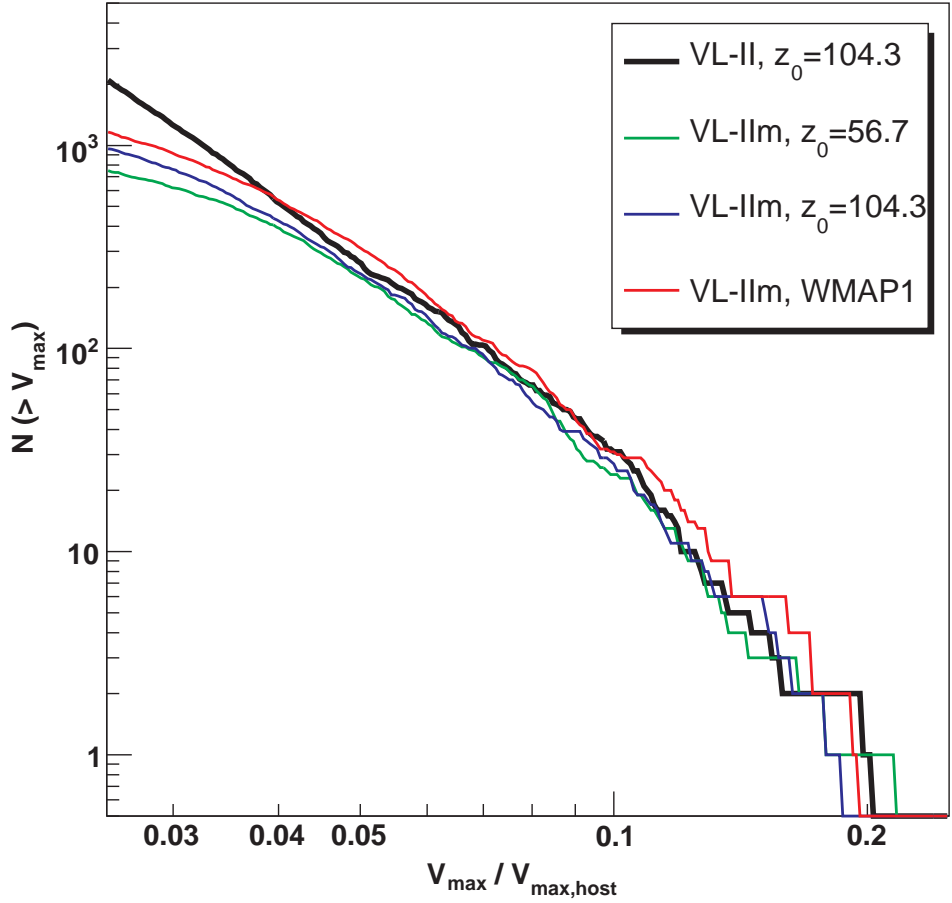}
\end{center}
\caption[]{Subhalo abundance at different numerical resolutions, starting redshifts and cosmologies. \rm
Number of subhalos above $\vmax / \vhost$ within $\rtwo$ for the VL-II simulation and three lower
resolution versions of the same halo.}
\label{fig:velfmVmaxS}
\end{figure}


\end{document}